\documentclass[prx,reprint,superscriptaddress,showpacs,amsmath,amssymb]{revtex4-1}

\usepackage[squaren]{SIunits}
\usepackage{graphicx}
\usepackage[latin1]{inputenc}

\usepackage{multirow}
\usepackage[linkcolor=blue,citecolor=blue,colorlinks=true,urlcolor=blue]{hyperref}
\renewcommand*{\vec}[1]{\mathbf{#1}}

\begin{document}


\title{\texorpdfstring{Independent ferroelectric contributions and rare-earth-induced polarization reversal in multiferroic TbMn$_2$O$_5$}{Independent ferroelectric contributions and rare-earth-induced polarization reversal in multiferroic TbMn2O5}}

\author{N.~Leo}
  \affiliation{Department of Materials, ETH Zurich, Wolfgang-Pauli-Strasse 10, 8093 Zurich, Switzerland}
  \affiliation{HISKP, Universit\"{a}t Bonn, Nussallee 14-16, 53115 Bonn, Germany}
\author{D.~Meier}
  \affiliation{Department of Physics, University of California, Berkeley, CA 94720, USA}
\author{R.V.~Pisarev}
  \affiliation{Ioffe Physical-Technical Institute, Russian Academy of Sciences, 194021 St.\ Petersburg, Russia}
\author{N.~Lee}
  \affiliation{Rutgers Center for Emergent Materials and Department of Physics \& Astronomy, Rutgers, The State University of New Jersey, Piscataway, NJ 08854, USA}
\author{S.-W.~Cheong}
  \affiliation{Rutgers Center for Emergent Materials and Department of Physics \& Astronomy, Rutgers, The State University of New Jersey, Piscataway, NJ 08854, USA}
\author{M.~Fiebig$^*$}
  \affiliation{Department of Materials, ETH Zurich, Wolfgang-Pauli-Strasse 10, 8093 Zurich, Switzerland}
  \affiliation{HISKP, Universit\"{a}t Bonn, Nussallee 14-16, 53115 Bonn, Germany}

\begin{abstract}
Three independent contributions to the magnetically induced spontaneous polarization of
multiferroic TbMn$_2$O$_5$ are uniquely separated by optical second harmonic generation and
an analysis in terms of Landau theory. Two of them are related to the magnetic Mn$^{3+/4+}$ order
and are independent of applied fields $\mu_0H_x$ of up to $\pm 7$~T. The third contribution is
related to the long-range antiferromagnetic Tb$^{3+}$ order. It shows a drastic decrease upon the
application of a magnetic field and mediates the change of sign of the spontaneous electric
polarization in TbMn$_2$O$_5$. The close relationship between the rare-earth long-range order and
the non-linear optical properties points to isotropic Tb$^{3+}$--Tb$^{3+}$ exchange and O$^{2-}$
spin polarization as mechanism for this rare-earth induced ferroelectricity.
\end{abstract}


\date{\today}

\maketitle


\section{Magnetoelectric multiferroics}\label{sec:introduction}

Materials with a coexistence of magnetic and electric long-range order, called multiferroics,
exhibit a variety of remarkable magnetoelectric cross-coupling phenomena. Ferroelectric phase
transitions driven by magnetic fields and magnetic phase transitions stimulated by electric fields
are among the most spectacular examples. A particularly rigid coupling between magnetic and
dielectric properties is obtained in multiferroics where the magnetic order induces an improper
ferroelectric polarization \cite{Newnham78,Kimura03a,Hur04Tb}. Such magnetically induced
ferroelectrics represent an outstanding class of multiferroics because they combine two desirable
aspects: (i) A unique control of spin-based properties by electric fields, and (ii) a matchless
versatility regarding intrinsic magnetoelectric coupling mechanisms. Prototypical examples are
TbMn$_2$O$_5$, TbMnO$_3$, and CuCrO$_2$ in which the spontaneous polarization is attributed to
magnetostrictive (exchange-symmetric), Dzyaloshinskii-Moriya-type (exchange-antisymmetric), and
orbital interactions, respectively \cite{Cheong07,Jia07,Arima07}.

A characteristic feature of the magnetically induced ferroelectrics is their complexity. Various
mechanisms, demanding in themselves, cooperate for promoting the spontaneous polarization:
geometric frustration, incommensurate spin order, low crystallographic symmetry, and electronic
$3d-4f$ interactions may all play a role. This inevitably results in rich magnetoelectric phase
diagrams with a multiplicity of contributions to the ferroelectric polarization $P$. The situation
culminates in compounds like TbMn$_2$O$_5$ which hosts three magnetic subsystems and five
different magnetically ordered phases. This leads to a complex temperature- and field-dependence
of the magnetically induced polarization, including a sign change of $P$ induced by a moderate
magnetic field \cite{Hur04Tb}. Phenomenological two- and three-polarization models were proposed to
explain such features and the $3d-4f$ interplay of the manganese and the terbium sublattices was
scrutinized \cite{Oh11,Johnson11}. However, up to now an explicit disentanglement of individual
contributions to the net polarization has not been accomplished because pyroelectric current
measurements, the standard technique for measuring values in the order of 1~nC/cm$^2$, can only
reveal the spatially integrated net polarization. However, a unique experimental separation of the
polarization contributions and their respective magnetic-field dependence would be highly valuable
for identifying the origin of magnetoelectric phase control in compounds like TbMn$_2$O$_5$.

Here we report that the spontaneous ferroelectric polarization in TbMn$_2$O$_5$ is composed of
\textit{three} independent contributions. Two of these are related to the magnetic order of the
Mn$^{3+/4+}$ ions while the third contribution originates in the antiferromagnetic Tb$^{3+}$
order, possibly involving isotropic exchange and O$^{2-}$ spin polarization by the rare-earth
order. We show that the magnetic-field-induced change of sign of the polarization in TbMn$_2$O$_5$
is dominated by the field-induced suppression of the Tb$^{3+}$-related polarization contribution (by to
field-induced quenching of the rare-earth order) and the emergence of an oppositely polarized
contribution (related to the quadratic magnetoelectric effect). In contrast, a response of the
manganese-related polarizations to the magnetic field is not observed, which is corroborated by
supplementary studies on isostructural YMn$_2$O$_5$. The sublattice polarizations in TbMn$_2$O$_5$
are disentangled by temperature- and magnetic-field dependent optical SHG and agree with a
multi-dimensional order parameter description by Landau theory.

It was initially proposed in Ref.~\onlinecite{Lottermoser09} that three independent contributions
might constitute the spontaneous ferroelectric polarization of TbMn$_2$O$_5$. However, the
separation was done in a purely phenomenological way. For demonstrating that a separation of
sublattice polarizations is possible in a unique way an analysis in terms of the Landau
formalism \cite{Toledano09,Menshenin09,Toledano87} is indispensable. In addition, attempts to
clarify the relation of the sublattice polarizations to the magnetically induced change of sign of
the polarization in TbMn$_2$O$_5$ were not made in Ref.~\onlinecite{Lottermoser09} where all
measurements were done in the absence of magnetic fields.

\begin{table}
\begin{tabular}{c||c|c|c}
 & $T_i$ & $\vec{k}$ & $P^y_\mathrm{net}$\\ \hline \hline
I & $T < T_1 = 43$~K & $(k_x,0,k_z)$ & -- \\ \hline
II & $T < T_2 = 38$~K & $(0.5,0,k_z)$ & \multirow{2}{*}{$P^y_I$} \\ \cline{1-3}
III & $T < T_3 = 33$~K & $(0.5,0,0.25)$ & \\ \hline
IV & $T < T_4 = 24$~K & \multirow{2}{*}{$(k_x,0,k_z)$} & $P^y_I - P^y_{II}$ \\ \cline{1-2}\cline{4-4}
V & $T<T_5 = 10$~K & & $P^y_I - P^y_{II} + P^y_{III}$
\end{tabular}
\caption{ Sequence of phase transitions in TbMn$_2$O$_5$ in zero magnetic field. $T_i$: transition
temperatures, $\vec{k}$: magnetic propagation vector, $P^y_\mathrm{net}$: magnetically induced net
polarization composed of the contributions $P^y_{I,II,III}$. Values for $T_i$ and $\vec{k}$ were
taken from Ref.~\onlinecite{Blake05}. } \label{tab:phases}
\end{table}

\section{\texorpdfstring{Multiferroic T\MakeLowercase{b}M\MakeLowercase{n}$_2$O$_5$}{Multiferroic TbMn2O5}}\label{sec:landau}

The isostructural \textit{R}Mn$_2$O$_5$ compounds (\textit{R}$^{3+}$ = Y, rare earth, Bi)
crystallize in the orthorhombic space group \textit{Pbam} (with the Cartesian $x$, $y$, and $z$ axis corresponding to the orthorhombic $a$, $b$, and $c$ axis, respectively).
Frustrated magnetic interactions between the Mn$^{3+}$ and Mn$^{4+}$ ions lead to commensurate and
incommensurate magnetic phases. In zero magnetic field, two successive transitions to an
antiferromagnetic (AFM) and to a multiferroic state, respectively, occur around 40~K in all compounds.
Additional magnetic transitions, in particular in compounds in which $R$ represents a rare-earth
element, were reported. In this Section we discuss the magnetic and multiferroic phases of
TbMn$_2$O$_5$ at zero magnetic field in detail and relate them to the corresponding description by
Landau theory in Ref.~\onlinecite{Toledano09}. The sequence of electric and magnetic phase
transitions is summarized in Table~\ref{tab:phases}.

In TbMn$_2$O$_5$, incommensurate AFM Mn$^{3+/4+}$ order in the $xy$ plane described
by the propagation vector $\vec{k}=(\frac{1}{2}+\delta_x,0,\frac{1}{4}+\delta_z)$ emerges at the
N\'{e}el temperature $T_N\equiv T_1=42$~K. The four-dimensional irreducible representation
associated with $\vec{k}$ leads to a ferroelectric polarization $P^y_\mathrm{net}$ along the $y$
axis when $\delta_x$ becomes zero in a second-order phase transition at $T_C\equiv T_2=38$~K.
Landau theory predicts a temperature dependence of the polarization according to
\begin{equation} \label{eq:P1a}
  P^y_\mathrm{net}(T_2>T>T_3) = \tilde{P}^y_I(T) \propto \sqrt{T_2-T} \, .
\end{equation}
In a subsequent second-order phase transition at $T_3=33$~K commensurate magnetic lock-in with
$\delta_x=\delta_z=0$ occurs. The predicted temperature dependence of the ferroelectric
polarization is now described by the expression
\begin{equation} \label{eq:P1b}
  P^y_\mathrm{net}(T_3>T>T_4) = P^y_I(T) \propto \sqrt{T_2-T+\varepsilon} \, .
\end{equation}
Here, $\varepsilon$ is a constant responsible for a change of slope of $P^y_\mathrm{net}(T)$ at
$T_3$. In a first-order phase transition at $T_4=22$~K the magnetic order of TbMn$_2$O$_5$ becomes
incommensurate again with $\delta_x\not=0$, $\delta_z\not=0$. The first-order phase transition is
parametrized by the emergence of now two four-dimensional order parameters whose coupling gives
rise to an additional polarization contribution $P^y_{II}$ with
\begin{equation}\label{eq:P2}
  P^y_{II}(T)\propto (T_4-T) \, .
\end{equation}
Because of the pronounced decrease of $P^y_\mathrm{net}$ observed at $T_4$ the net polarization is
expressed as
\begin{equation} \label{eq:Pnet}
  P^y_\mathrm{net}(T<T_4) = P^y_I(T) - P^y_{II}(T) = \varepsilon_1 \sqrt{T_2-T + \varepsilon} - \varepsilon_2(T_4-T)
\end{equation}
with $\varepsilon_{1,2}$ as proportionality factors. Finally, a second-order magnetic phase
transition is observed at $T_5\approx 10$~K. The transition is associated to the long-range
AFM order of the Tb$^{3+}$ moments with the same propagation vector as for the
Mn$^{3+/4+}$ moments \cite{Kobayashi04Tb,Johnson08}. It is accompanied by a recovery of the
ferroelectric net polarization. Hence, a dome-like temperature dependence of $P^y_{II}$ or,
alternatively, a third contribution $P^y_{III}$ to the net polarization were
proposed \cite{Hur04Tb,Lottermoser09}.

\section{Optical second harmonic generation (SHG)}\label{sec:shg}

Optical SHG describes the induction of a light wave at frequency $2\omega$ by an incident light
wave at frequency $\omega$ \cite{Shen_nlo,Fiebig05b}. This is expressed as
$S_i(2\omega)=\epsilon_0\chi_{ijk}E_j(\omega)E_k(\omega)$. The component $\chi_{ijk}$ of the
corresponding non-linear susceptibility tensor couples $j$ and $k$ polarized contributions of the
incident light field $\vec{E}(\omega)$ to an $i$ polarized contribution of the SHG source term
$\vec{S}(2\omega)$. The according SHG intensity is $I_{ijk}\propto|\vec{S}(2\omega)|^2\propto
|\chi_{ijk}|^2$. In the electric-dipole approximation $\hat{\chi}$ is a polar tensor with
$\hat{\chi}\neq 0$ in non-centrosymmetric systems only \cite{Shen_nlo}. Thus, SHG is well suited
for detecting ferroelectric order breaking the inversion symmetry \cite{Uesu95}. Moreover, contrary
to linear optical techniques, the ferroelectric SHG contribution emerges free of background. In
addition, the polarization of the light waves at $\omega$ and $2\omega$ that are involved in the
SHG process reveals the symmetry and orientation of the ferroelectric order parameter, i.e., the
direction of the spontaneous polarization.

An analysis of the spectral dependence of the SHG signal reveals the electronic states
contributing to the local electric-dipole moment and, hence, to the macroscopic polarization. Thus,
independent contributions to the ferroelectric polarization can be separated by measurements of
the SHG signal in dependence of the polarization of the light fields, photon energy, temperature and magnetic field.

\section{Experimental setup}\label{sec:setup}

SHG was measured with 5-ns laser pulses using the transmission setup described
elsewhere \cite{Fiebig05b,Lottermoser09}. We used (100)-oriented TbMn$_2$O$_5$ and YMn$_2$O$_5$
samples with a thickness of 50 and 100~$\mu$m, respectively, which were polished with a silica
slurry. The samples were mounted in a temperature-tunable liquid-helium-operated cryostat for
fields up to 7~T (Oxford Spectromag SM4000). The magnetic field was applied in the Faraday
configuration ($\vec{H}\,\|\,x$). The integrated SHG intensity of the sample was measured while
heating it (heat rate $0.5-1$~K/min) or sweeping the magnetic field (sweep rate $0.25-0.5$~T/min).
An electric single-domain poling procedure was omitted because even in zero electric field the
samples were found to predominantly form a ferroelectric single-domain state.

\begin{figure}[bt]
\includegraphics[width=8.6cm]{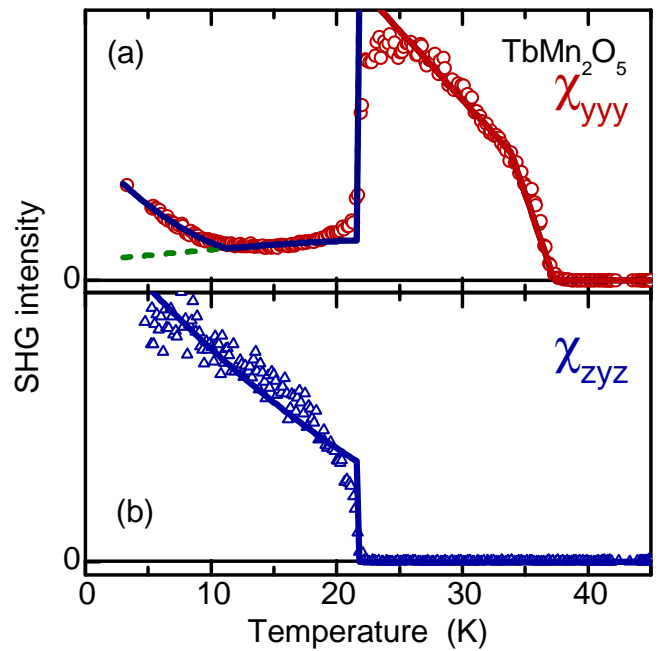}
 \caption{
  (Color online)
Separation of sublattice polarizations in TbMn$_2$O$_5$. Temperature dependence of the SHG
contributions $I_{ijk}\propto|\chi_{ijk}|^2$ for (a) $\chi_{yyy}$ and (b) $\chi_{zyz}$ measured in
the same experimental run at $2\hbar\omega$=2.08~eV in zero magnetic field. Three independent
contributions $P^y_{I,II,III}$ to the net polarization are uniquely distinguished (see text).
Whereas $\chi_{yyy}$ in (a) couples to a coherent superposition of all three sublattice
polarizations, $\chi_{zyz}$ in figure (b) couples exclusively to the polarization $P^y_{II}(T)$.
Lines are fits in agreement with Landau theory (Eqs.~\protect{\ref{eq:P1a}} to
\protect{\ref{eq:fullmodel}}) as explained in Section~\ref{sec:h-zero}. Note that the data below 10~K
in (a) cannot be explained by a sum of $P^y_{I}$ and $P^y_{II}$ (dashed line), so that a third
contribution, $P^y_{III}$, related to the magnetic order of the Tb$^{3+}$ ions has to be taken
into account (solid line). }
 \label{im:Tb_zero_field}
\end{figure}

\section{Contributions to the zero-field net polarization}\label{sec:h-zero}

We first present measurements at zero magnetic field in order to demonstrate the method of
separating independent contributions to the ferroelectric net polarization and verify the
consistency between the temperature dependence of these contributions and the behaviour expected
from Landau theory.

Figure~\ref{im:Tb_zero_field} shows the temperature dependence of the SHG signals $I_{yyy}$
and $I_{zyz}$ in TbMn$_2$O$_5$ at $2\hbar\omega$=2.08~eV. These non-linear intensities were
chosen because they allow a unique distinction of polarization contributions \cite{Lottermoser09}:
SHG from $\chi_{yyy}$ is present below $T_2\approx 38$~K which marks the emergence of the
magnetically induced polarization. Note that the temperature dependence of
$\sqrt{I_{yyy}}\propto|\chi_{yyy}|$ resembles pyroelectric current measurements of the net
polarization \cite{Hur04Tb,Cruz07}. In contrast, $I_{zyz}$ is non-zero only below the first-order
commensurate-to-incommensurate transition at $T_4\approx 22$~K. Thus, at least two independent
contributions to the net polarization are present which correspond to separate SHG light fields
interfering in a different way in Figs.~\ref{im:Tb_zero_field}(a) and \ref{im:Tb_zero_field}(b).

We now apply a systematic fit procedure employing the temperature dependence as predicted by Landau
theory \cite{Toledano09} in Section~\ref{sec:landau}. For $T_2\ge T>T_4$ the
intensity $I_{yyy}$ depends only on the polarization $P^y_I(T)$ according to Eqs.~(\ref{eq:P1a})
and (\ref{eq:P1b}) in good agreement with the fit in Fig.~\ref{im:Tb_zero_field}(a). For $T_4\ge
T\ge T_5$ the additional polarization $P^y_{II}$ is present. In this temperature range $I_{zyz}$
in Fig.~\ref{im:Tb_zero_field}(b) couples solely to $P^y_{II}$ and reveals the relation
$I_{zyz}\propto T^2$ expected from Eq.~(\ref{eq:P2}). In contrast, $I_{yyy}$ includes interfering
SHG contributions from $P^y_I$ and $P^y_{II}$. This is described by
\begin{equation}\label{eq:shortmodel}
  I_{yyy}(T_4 \ge T \ge T_5) = I_0|P^y_I(T) + Ae^{i\phi}P^y_{II}(T)|^2
\end{equation}
Here, $P^y_{I,II}(T)$ are derived by fitting Eqs.~(\ref{eq:P1a}) to (\ref{eq:P2}), respectively,
in the temperature range where they constitute the only contribution to the SHG signal: (i)
$P^y_I(T)$ is derived by fitting Eqs.~(\ref{eq:P1a}) and (\ref{eq:P1b}) to $I_{yyy}(T)$ in the
range $T_4 < T \le T_2$. (ii) $P^y_{II}(T)$ is derived by fitting Eq.~(\ref{eq:P2}) to
$I_{zyz}(T)$ for $T\le T_4$. For fitting Eq.~(\ref{eq:shortmodel}) to $I_{yyy}(T\le T_4)$, where
$P^y_I$ and $P^y_{II}$ interfere, the relative amplitude $A$ and the phase $\phi$ were varied
whereas $P^y_{I,II}$ were adopted from the fits in (i) and (ii). For $T\ge T_5$ the fits leads to
a good agreement between data and theory in Fig.~\ref{im:Tb_zero_field} except near $T_4$. Here,
the change of the SHG intensity is smeared out due to the abrupt nature of the first-order
transition and because of a thermal gradient across the sample caused by laser heating. This also
leads to small deviation between the transition temperatures $T_{2,3,4}$ obtained here and their
literature values \cite{Hur04Tb,Radaelli08}.

Below $T_5\approx 10$~K Eq.~(\ref{eq:shortmodel}) fails to describe the temperature dependence of
the SHG data in Fig.~\ref{im:Tb_zero_field}(a) which indicates the presence of a \textit{third}
contribution to the net polarization of TbMn$_2$O$_5$. The value of $T_5$ shows that the
long-range order of the Tb$^{3+}$ lattice is responsible for this contribution. Hence, the
corresponding ferroelectric polarization $P^y_{III}$ is improper so that the temperature dependence
has to scale linearly according to Landau theory:
\begin{equation}\label{eq:P3}
  P^y_{III}(T)\propto (T_5-T)\, .
\end{equation}
In an extension of Eq.~(\ref{eq:shortmodel}) the corresponding net SHG intensity is described by
\begin{equation}\label{eq:fullmodel}
  I_{yyy}(T<T_5) = I_0|P^y_I(T) + Ae^{i\phi}P^y_{II}(T) + A^{\prime}e^{i\phi^\prime}P^y_{III}(T)|^2 \, .
\end{equation}
For this fit only the amplitude $A^{\prime}$ and phase $\phi^\prime$ were varied whereas all the
other values were taken from the fit of Eq.~(\ref{eq:shortmodel}) for $T_4\ge T \ge T_5$. As
Fig.~\ref{im:Tb_zero_field} shows, this leads to an excellent agreement with the SHG data down to
the lowest experimentally accessed temperature of about 3~K.

We thus conclude that SHG allows us to decompose the spontaneous ferroelectric net polarization of
TbMn$_2$O$_5$ in a unique way into three independent sublattice polarizations $P^y_{I,II,III}$.
All three contributions couple linearly to the SHG susceptibility and exhibit the temperature
dependence that is expected from Landau theory. The critical temperatures at which the sublattice
polarizations emerge relate $P^y_I$ and $P^y_{II}$ to the magnetic order of the Mn$^{3+/4+}$
lattices and the ``new'' contribution $P^y_{III}$ to the AFM order of the Tb$^{3+}$ lattice. The
macroscopic net polarization is conveniently written as
\begin{equation}\label{eq:net_polarization}
  P^y_\mathrm{net}(T) = P^y_I(T) - P^y_{II}(T) + P^y_{III}(T)\, .
\end{equation}
This reflects the antiparallel orientation of $P^y_{I}(T)$ and $P^y_{II}(T)$ which is responsible
for the pronounced decrease of $P^y_\mathrm{net}(T)$ at $T_4$.

\begin{figure}[bt]
\includegraphics[width=8.6cm]{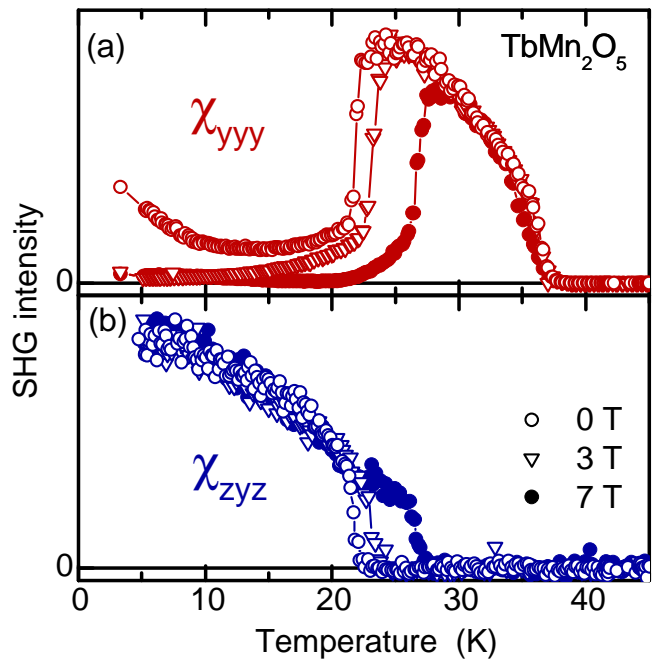}
 \caption{
  (Color online)
Temperature dependence of the SHG contributions from (a) $\chi_{yyy}$ and (b) $\chi_{zyz}$
measured in the same experimental run at $2\hbar\omega$=2.08~eV at different magnetic fields
$\mu_0H_x$. Note the suppression of $\chi_{yyy}$ at low temperatures~(a) and the shift of the
first-order transition temperature from 22~K to 27~K~(a,b). In spite of this shift the amplitude
of neither $\chi_{yyy}$ above 30~K (in (a)) nor of $\chi_{zyz}$ (in (b)) show a magnetic-field
dependence. }
 \label{im:Tb_mag_field}
\end{figure}

\begin{figure}[bt]
\includegraphics[width=8.6cm]{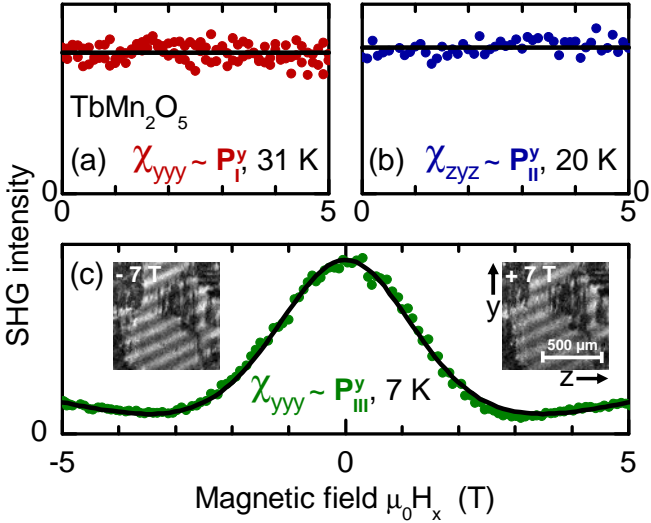}
 \caption{
  (Color online)
Magnetic field dependence of the sublattice polarizations of TbMn$_2$O$_5$: SHG from (a)
$\chi_{yyy}$ at 31~K and (b) $\chi_{zyz}$ at 20~K for $2\hbar\omega$=2.08~eV while sweeping the
magnetic field $\mu_0H_x$. These contributions couple to the Mn$^{3+/4+}$-related polarizations
$P^y_I$ and $P^y_{II}$, respectively, and do not display a magnetic-field dependence. (c) SHG from
$\chi_{yyy}$ at low temperatures. This contribution couples to the Tb$^{3+}$-related polarization
$P^y_{III}$ and displays a pronounced magnetization dependence. The solid line represents a fit of
Eq.~(\ref{eq:brillouin}) to the data. Insets in (c) show that the ferroelectric domain structure is independent of the applied magnetic field. Insets show spatially resolved SHG images of TbMn$_2$O$_5$ for $\chi_{yyy}$ at 2.08~eV and 5~K in fields of $-7$~T and $+7$~T, respectively. Note that the position of the ferroelectric domains and domain walls (dark regions) does not change in the course of the magnetic field reversal. (The periodic overall variation of brightness are interference fringes of the fundamental light in the sample.)}
 \label{im:Tb_field_comparison}
\end{figure}

\begin{figure}
 \includegraphics[width=8.6cm]{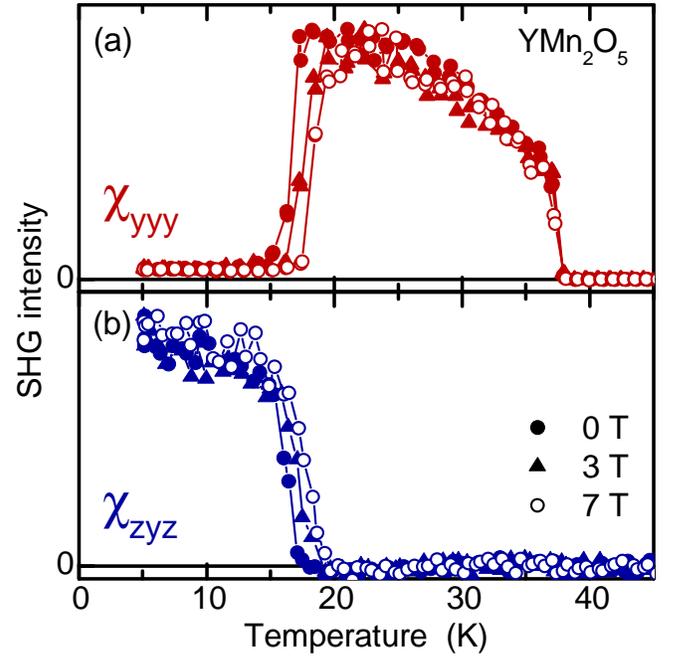}
 \caption{
  (Color online)
Magnetic-field dependence of sublattice polarizations in YMn$_2$O$_5$. (a,b) Temperature
dependence of the SHG contributions from (a) $\chi_{yyy}$ and (b) $\chi_{zyz}$ measured in the
same experimental run at $2\hbar\omega$=2.08~eV at different magnetic fields $\mu_0H_x$. Like in
TbMn$_2$O$_5$, two contributions to the polarization, $P^y_I$ and $P^y_{II}$, can be
distinguished. However, in contrast to Fig.~\ref{im:Tb_mag_field} the rare-earth
contribution below 10~K is absent. }
 \label{im:Y_mag_field}
\end{figure}

\section{Magnetic-field dependence of the polarization}\label{sec:h-field}

After establishing the relation between the SHG signal and the ferroelectric net polarization in
TbMn$_2$O$_5$ we now investigate the response of the SHG yield to a static magnetic field applied
along the $x$ axis. Figures~\ref{im:Tb_mag_field}(a) and \ref{im:Tb_mag_field}(b) show the
temperature dependence of $I_{yyy}$ and $I_{zyz}$ for magnetic fields $\mu_0H_x$ of 0, 3, and 7~T.
The most pronounced effects of the magnetic field are the shift \cite{Hur04Tb,Radaelli08} of the
first-order transition temperature $T_4$ and the suppression of $I_{yyy}$ below this temperature.

We now apply the same step-by-step analysis as in Section~\ref{sec:h-zero} with consecutive
derivation of $P^y_I$, $P^y_{II}$, and $P^y_{III}$ and the associated fit parameters according to
Eq.~(\ref{eq:fullmodel}). For this purpose, Fig.~\ref{im:Tb_field_comparison} shows the
magnetic-field dependence of $I_{yyy}$ and $I_{zyz}$ at fixed photon energies and temperatures.
The SHG signals in Figs.~\ref{im:Tb_field_comparison}(a) and \ref{im:Tb_field_comparison}(b) are
constant across the whole range of the magnetic-field. As shown in Section~\ref{sec:h-zero},
$I_{yyy}$ at 30~K and 2.08~eV couples solely to $P^y_I$ whereas $I_{zyz}$ at 20~K and 2.08~eV is
only determined by $P^y_{II}$. We therefore conclude that, although the commensurate phase is
notably stabilized by the magnetic field (see the shift of the first-order-transition temperature in
Fig.~\ref{im:Tb_mag_field}(a) and Refs.~\onlinecite{Hur04Tb} and \onlinecite{Radaelli08}), the
absolute values of the polarizations $P^y_I$ and $P^y_{II}$ are \textit{not affected} by fields up
to $\pm 7$~T. Consequently, the pronounced magnetoelectric response in TbMn$_2$O$_5$ must be
\textit{solely} due to the rare-earth induced polarization contribution. This is corroborated in
an impressive way by SHG data on YMn$_2$O$_5$ which are shown in Fig.~\ref{im:Y_mag_field}: The
temperature dependence of $I_{yyy}$ and $I_{zyz}$ is similar to that of TbMn$_2$O$_5$ and allows
us to distinguish the sublattice polarizations $P^y_{I}$ and $P^y_{II}$ according to
Eqs.~(\ref{eq:P1a}) to (\ref{eq:shortmodel}). However, contributions from $P^y_{III}$ are absent
because of the diamagnetic Y$^{3+}$ sublattice. Along with this, the data in
Fig.~\ref{im:Y_mag_field} display no dependence on the applied magnetic field aside from a small
shift of the first-order-transition temperature $T_4$. The role of $P^y_{III}$ in TbMn$_2$O$_5$ is
highlighted in Fig.~\ref{im:Tb_field_comparison}(c) which shows $I_{yyy}(H_x)$ at 7~K and 2.08~eV.
The SHG signal displays minima at $\pm 3$~T and a steady increase of the SHG intensity away from
this value. Because of the independence of $P^y_{I,II}$ on magnetic field
Fig.~\ref{im:Tb_field_comparison}(c) directly reflects the magnetic-field dependence of
$P^y_{III}$. The symmetric shape of $I_{yyy}$ reveals that there is no hysteresis. In addition,
the response is independent of the sign of $H_x$ which indicates a quadratic field-dependence of
$P^y_{III}(H_x)$. An expression taking these observations into account is
\begin{equation}\label{eq:brillouin}
  P^y_{III}(T,H_x) = P^y_{III}(T,0) - \beta M_x^2(T,H_x) \, .
\end{equation}
Here we assume that, as discussed below, the alignment of the large magnetic moment of the
Tb$^{3+} (4f^8)$ ions leads to a magnetization described by a Brillouin function according to
$M_x\propto B_J(T,H_x)$ with $J=6$ and $g_J=1.5$ for Tb$^{3+}$. The data in
Fig.~\ref{im:Tb_field_comparison}(c) can now be fitted by Eq.~(\ref{eq:fullmodel}) into which we
enter $P^y_{III}(H_x)$ according to Eq.~(\ref{eq:brillouin}) as well as the field independent
polarizations $P^y_{I,II}$. We see that the agreement between the SHG data and the fit is
excellent.

\begin{figure}
 \includegraphics[width=8.6cm]{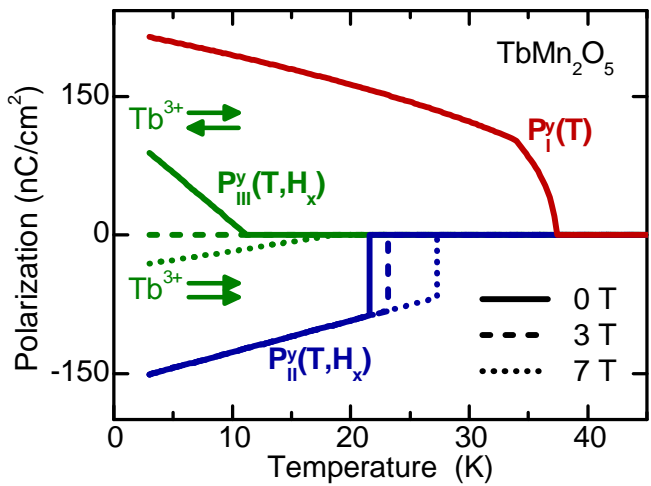}
 \caption{
 (Color online)
Temperature- and magnetic-field-dependent decomposition of the magnetically induced polarization $P_\mathrm{net}\,\|\,y$
in TbMn$_2$O$_5$. The polarizations $P^y_I$ and $P^y_{II}$ induced by the manganese spin order are
field-independent (apart from the shift of the first-order-transition temperature $T_4$). These
contributions are also present in YMn$_2$O$_5$ (see Fig.~\ref{im:Y_mag_field}). The
Tb$^{3+}$-related polarization contribution $P^y_{III}$ in zero field is induced by the AFM
long-range order of the rare-earth moments.
In an applied magnetic field $H||x$ this contribution is suppressed due to the paramagnetic magnetization of the Tb$^{3+}$ moments whereas another field-induced contribution (related to the quadratic magnetoelectric effect) emerges. Both effects lead to the magnetic-field induced sign change of $P_\mathrm{net}$ as observed in Ref.~\onlinecite{Hur04Tb}.
}
 \label{im:3pol_model}
\end{figure}

Figure~\ref{im:3pol_model} shows the resulting field- and temperature-dependent
decomposition of $P^y_\mathrm{net}(T,H_x)$. The magnetic-field-induced change of sign of the
polarization at low temperatures \cite{Hur04Tb} reveals itself as the suppression and zero crossing
of a single contribution, $P^y_{III}$, to the net polarization. Therefore, this process should not
be termed ``polarization reversal'' or ``polarization switching'', because these terms are
associated to the change of polarization between $+P$ and $-P$ involving a hysteresis with the
formation and movement of domains. In TbMn$_2$O$_5$ only the balance between competing contributions to the
net polarization is shifted whereas the domain structure reveals small or no changes according to
our SHG imaging experiments (see Fig.~\ref{im:Tb_field_comparison}~(c)).

\section{Discussion: Microscopic mechanisms}\label{sec:discussion}

The essential benefit of the SHG experiments is that they allow us to separate independent
contributions to the magnetically induced ferroelectric polarization in TbMn$_2$O$_5$ in a unique
way and investigate their respective response to a static magnetic field. As we will see now, this
leads to clues about the microscopic origin of these polarizations and of the magnetically induced
change of sign of the polarization.

Experimentally we distinguish three contributions to the net polarization: The first two,
$P^y_{I}$ and $P^y_{II}$, are associated to the magnetic order of the Mn$^{3+/4+}$ sublattices.
They maintain an antiparallel orientation, leading to a sudden drop of the net polarization once
$P^y_{II}$ appears at $T_4$. This was assumed before but here the separate observation of $P^y_{I}$
and $P^y_{II}$ provides an unambiguous confirmation of their opposite sign. In addition, the
observation of two independent contributions rules out the model explaining the polarization drop
at $T_4$ as transition from a ferroelectric state for $T<T_4$ to a ferrielectric state for
temperatures above $T_4$ \cite{Fukunaga10}. The magnitudes of $P^y_{I}$ and $P^y_{II}$ respond
neither to magnetic fields $\mu_0H_x$ of up to $\pm 7$~T nor to the magnetic order of the
Tb$^{3+}$ sublattice \cite{Radaelli08}. For $P^y_{I}$ this has been seen previously in the
restricted temperature range $T>T_4$ where only $P^y_{I}$ contributes to the net polarization. Now
the SHG data in Fig.~\ref{im:Tb_mag_field} give direct access to $P^y_{I}$ and $P^y_{II}$ across
the entire temperature range. In contrast to field and temperature dependent pyroelectric current
measurements \cite{Oh11}, which revealed changes in the order of 10\%/T, the field dependent SHG
data of $P^y_{I,II}$ at fixed temperature (Figs.~\ref{im:Tb_field_comparison}(a) and (b)) reveal
no dependence on the magnetic field (with a noise level of the SHG data below 1\%). It is quite
remarkable that the absolute values of $P^y_I$ and $P^y_{II}$ are field independent although the
magnetic field stabilizes the incommensurate phase ($\sim P^y_{II}$) against the commensurate
phase ($\sim P^y_I$) at $T_4$.

The third contribution to the net polarization, $P^y_{III}$, is associated to the magnetic order
of the Tb$^{3+}$ ions. This is the most interesting contribution because it is the only one
displaying a magnetic-field dependence and is therefore exclusively responsible for the
magnetic-field-induced change of sign of the net polarization \cite{Hur04Tb}. First of all, we see
that the emergence of $P^y_{III}$ in zero magnetic field is associated to a \textit{specific
component} of the magnetic Tb$^{3+}$ order. We have to distinguish (i) the magnetic order exerted
onto the Tb$^{3+}$ ions by the ordered Mn$^{3+/4+}$ spins, and (ii) the magnetic order resulting
from isotropic Tb$^{3+}$--Tb$^{3+}$ exchange \cite{Wehrenfennig10}. Contribution (i) is observed in
the whole temperature range below $T_1$ \cite{Blake05,Koo07,Johnson08}. Contribution (ii) is
present below $T_5$ and distinguished from contribution (i) by a small characteristic hump in the
dielectric resonance \cite{Kobayashi04Tb,Cruz07}. The emergence of $P^y_{III}$ at $T_5$ clearly
relates it to the contribution (ii): the AFM order inherent to the Tb$^{3+}$ sublattice.
$P^y_{III}$ is linear in the temperature which matches the increase of ordered magnetic moment of
the Tb$^{3+}$ ions below 10~K \cite{Chapon04,Kobayashi04Tb,Koo07,Johnson08}.

The field dependence of the magnetic Tb$^{3+}$ order and of $P^y_{III}(H_x)$, expressed by
Fig.~\ref{im:Tb_field_comparison} and Eq.~(\ref{eq:brillouin}), points to the competition of two
independent mechanisms to the magnetoelectric response. On the one hand, we have
$P^y_{III,\mathrm{AFM}}(H_x)$, which is related to the long-range ordered AFM component of the
rare-earth sublattice. This is the only contribution present at $H_x=0$. At $H_x\neq 0$ this
polarization is \textit{diminished} by the fraction of the magnetic rare-earth moment that is
paramagnetically aligned along $x$ \cite{Johnson11}. This is expressed by
$P^y_{III,\mathrm{AFM}}(H_x)=P^y_{III,\mathrm{AFM}}(0)-\beta_1M_x^2$ with $M_x$ as paramagnetic
magnetization and $\beta_1$ determined by $\lim_{H_x\to\infty}P^y_{III,\mathrm{AFM}}(H_x)=0$. On
the other hand, we have $P^y_{III,\mathrm{PM}}(H_x)$, which is the polarization \textit{induced}
by the paramagnetically aligned component of the rare-earth moments. This is expressed by
$P^y_{III,\mathrm{PM}}(H_x)=\beta_2M_x^2$ with $\beta_2$ as susceptibility parametrizing this
quadratic magnetoelectric effect \cite{Saito95,Nakamura97}.

We note that $P^y_{III,\mathrm{AFM}}$ is due to the internal exchange interactions involving a scalar product of the Tb$^{3+}$ moments, whereas $P^y_{III,\mathrm{PM}}(H_x)$ is related to the product of the magnetic moment vector and the external magnetic field inducing $M_x$.

Since both the AFM demagnetization and
the quadratic magnetoelectric effect depend quadratically on the paramagnetic magnetization, the
two contributions are jointly described by the coefficient $\beta=\beta_1+\beta_2$ in
Eq.~(\ref{eq:brillouin}). The paramagnetic behaviour is reflected by the Brillouin function fitting
$M_x(H_x)$ in Fig.~\ref{im:Tb_field_comparison} \cite{Saito95,Hur04Tb}. Note that beyond about $\pm
3$~T the quadratic magnetoelectric contribution exceeds the remaining AFM contribution to the
polarization so that the change of sign of $P^y_{III}$ occurs. Furthermore, $P^y_{III,PM}(H_x)$
persists up to temperatures $T>T_5$ \cite{Hur04Tb,Oh11}, which once more emphasizes its relation to
the paramagnetic magnetization (in contrast to the AFM long-range Tb$^{3+}$ order disappearing at $T_5$).

The quadratic dependence of $P^y_{III}$ on the magnetic field in Eq.~(\ref{eq:brillouin}) can be
understood in two ways. Macroscopically, it expresses that the magnetoelectric polarization does
not depend on the sign of $H_x$, which is also in agreement with symmetry considerations.
Microscopically, it expresses the relation of the magnetoelectric polarization to the isotropic
Tb$^{3+}$--Tb$^{3+}$ exchange.

The SHG data reveal further clues about the microscopic origin of $P^y_{III}$: On the one hand,
SHG probes the $\mathrm{Mn}\to\mathrm{O}$ charge-transfer
excitation \cite{Moskvin08Tb,Lottermoser09} and can thus be sensitive to the magnetic Mn$^{3+/4+}$
order and the related polarization $P^y_{I,II}$. On the other hand, SHG couples to the
Tb$^{3+}$-related polarization $P^y_{III}$ although optical Tb$^{3+}$ excitations are not present
in the range where the SHG data were taken. This relates $P^y_{III}$ to an exchange interaction
between the Tb$^{3+}$ and the Mn$^{3+/4+}$  spins. However the SHG data \textit{do not} simply
reproduce the magnetic moment exerted onto the Mn$^{3+/4+}$ ions by the ordered Tb$^{3+}$ spins,
as this would also affect $P^y_{I}$ and $P^y_{II}$. This inevitably leads to the oxygen as only
other ingredient involved in the Tb$^{3+}$--Mn$^{3+/4+}$ exchange coupling. We thus propose that
$P^y_{III}$ originates in O$^{2-}$ ions that are spin-polarized by the Tb$^{3+}$ long-range order
and therefore affect the $\mathrm{Mn}\to\mathrm{O}$ charge transfer probed by SHG. Recent results
showing the importance of spin-polarized oxygen for the ferroelectric properties of
\textit{R}Mn$_2$O$_5$ support this conclusion \cite{Beale10,Partzsch11}.

\section{Conclusions}\label{sec:conclusions}

Three independent contributions to the magnetically induced spontaneous polarization of
multiferroic TbMn$_2$O$_5$ were separated by SHG. All three display a temperature-dependent
behaviour in accordance with Landau theory. Two contributions, $P^y_I$ and $P^y_{II}$, are related
to the magnetic Mn$^{3+/4+}$ order; they were also observed in YMn$_2$O$_5$ without rare-earth
magnetism. They are insensitive to a magnetic field $\mu_0H_x$ of up to $\pm 7$~T within 1\% of
the maximal polarization value. The third contribution, $P^y_{III}$, is induced by the AFM
long-range order of the Tb$^{3+}$ moments, as deduced from the observed temperature and
magnetic-field dependence of $P^y_{III}(T,H_x)$. Suppression of this rare-earth order and the
emergence of a paramagnetic magnetization in a magnetic field $H_x$ lead to the change of sign of
$P^y_{III}$ in TbMn$_2$O$_5$. Our data reveal isotropic Tb$^{3+}$--Tb$^{3+}$ exchange and spin
polarized oxygen as likely mechanism behind the rare-earth-induced ferroelectric contribution.

\begin{acknowledgments}
The authors thank K.~Kohn and V.A.~Sanina for providing the TbMn$_2$O$_5$ single crystals
and P.~Tol\'edano for many fruitful discussions. Financial support by the SFB~608 of the DFG is
appreciated. R.V.P. acknowledges support by the RFBR project 09-02-00070. The work at Rutgers was
supported by National Science Foundation DMR-1104484.
\end{acknowledgments}


$^\ast$ manfred.fiebig@mat.ethz.ch


\begin{thebibliography}{31}
\expandafter\ifx\csname natexlab\endcsname\relax\def\natexlab#1{#1}\fi
\expandafter\ifx\csname bibnamefont\endcsname\relax
  \def\bibnamefont#1{#1}\fi
\expandafter\ifx\csname bibfnamefont\endcsname\relax
  \def\bibfnamefont#1{#1}\fi
\expandafter\ifx\csname citenamefont\endcsname\relax
  \def\citenamefont#1{#1}\fi
\expandafter\ifx\csname url\endcsname\relax
  \def\url#1{\texttt{#1}}\fi
\expandafter\ifx\csname urlprefix\endcsname\relax\def\urlprefix{URL }\fi
\providecommand{\bibinfo}[2]{#2}
\providecommand{\eprint}[2][]{\url{#2}}

\bibitem[{\citenamefont{Newnham et~al.}(1978)\citenamefont{Newnham, Kramer,
  Schulze, and Cross}}]{Newnham78}
\bibinfo{author}{\bibfnamefont{R.~E.} \bibnamefont{Newnham}},
  \bibinfo{author}{\bibfnamefont{J.~J.} \bibnamefont{Kramer}},
  \bibinfo{author}{\bibfnamefont{W.~A.} \bibnamefont{Schulze}},
  \bibnamefont{and} \bibinfo{author}{\bibfnamefont{L.~E.} \bibnamefont{Cross}},
  \bibinfo{title}{\textit{Magnetoferroelectricity in Cr$_2$BeO$_4$}},
  \bibinfo{journal}{J. Appl. Phys.} \textbf{\bibinfo{volume}{49}},
  \bibinfo{pages}{6088 } (\bibinfo{year}{1978}).

\bibitem[{\citenamefont{Kimura et~al.}(2003)\citenamefont{Kimura, Goto,
  Shintani, Ishizaka, Arima, and Tokura}}]{Kimura03a}
\bibinfo{author}{\bibfnamefont{T.}~\bibnamefont{Kimura}},
  \bibinfo{author}{\bibfnamefont{T.}~\bibnamefont{Goto}},
  \bibinfo{author}{\bibfnamefont{H.}~\bibnamefont{Shintani}},
  \bibinfo{author}{\bibfnamefont{K.}~\bibnamefont{Ishizaka}},
  \bibinfo{author}{\bibfnamefont{T.}~\bibnamefont{Arima}}, \bibnamefont{and}
  \bibinfo{author}{\bibfnamefont{Y.}~\bibnamefont{Tokura}},
  \bibinfo{title}{\textit{Magnetic control of ferroelectric polarization}},
  \bibinfo{journal}{Nature (London)} \textbf{\bibinfo{volume}{426}}, \bibinfo{pages}{55}
  (\bibinfo{year}{2003}).

\bibitem[{\citenamefont{Hur et~al.}(2004)\citenamefont{Hur, Park, Sharma, Ahn,
  Guha, and Cheong}}]{Hur04Tb}
\bibinfo{author}{\bibfnamefont{N.}~\bibnamefont{Hur}},
  \bibinfo{author}{\bibfnamefont{S.}~\bibnamefont{Park}},
  \bibinfo{author}{\bibfnamefont{P.}~\bibnamefont{Sharma}},
  \bibinfo{author}{\bibfnamefont{J.}~\bibnamefont{Ahn}},
  \bibinfo{author}{\bibfnamefont{S.}~\bibnamefont{Guha}}, \bibnamefont{and}
  \bibinfo{author}{\bibfnamefont{S.-W.} \bibnamefont{Cheong}},
  \bibinfo{title}{\textit{Electric polarization reversal and memory in a multiferroic material induced by magnetic fields}},
  \bibinfo{journal}{Nature (London)} \textbf{\bibinfo{volume}{429}},
  \bibinfo{pages}{392} (\bibinfo{year}{2004}).

\bibitem[{\citenamefont{Cheong and Mostovoy}(2007)}]{Cheong07}
\bibinfo{author}{\bibfnamefont{S.-W.} \bibnamefont{Cheong}} \bibnamefont{and}
  \bibinfo{author}{\bibfnamefont{M.}~\bibnamefont{Mostovoy}},
  \bibinfo{title}{\textit{Multiferroics: a magnetic twist for ferroelectricity}},
  \bibinfo{journal}{Nature Mater.} \textbf{\bibinfo{volume}{6}},
  \bibinfo{pages}{13} (\bibinfo{year}{2007}).

\bibitem[{\citenamefont{Jia et~al.}(2007)\citenamefont{Jia, Onoda, Nagaosa, and
  Han}}]{Jia07}
\bibinfo{author}{\bibfnamefont{C.}~\bibnamefont{Jia}},
  \bibinfo{author}{\bibfnamefont{S.}~\bibnamefont{Onoda}},
  \bibinfo{author}{\bibfnamefont{N.}~\bibnamefont{Nagaosa}}, \bibnamefont{and}
  \bibinfo{author}{\bibfnamefont{J.~H.} \bibnamefont{Han}},
  \bibinfo{title}{\textit{Microscopic theory of spin-polarization coupling in multiferroic transition metal oxides}},
  \bibinfo{journal}{Phys. Rev. B} \textbf{\bibinfo{volume}{76}},
  \bibinfo{pages}{144424} (\bibinfo{year}{2007}).

\bibitem[{\citenamefont{Arima}(2007)}]{Arima07}
\bibinfo{author}{\bibfnamefont{T.}~\bibnamefont{Arima}},
  \bibinfo{title}{\textit{Ferroelectricity Induced by Proper-Screw Type Magnetic Order}},
  \bibinfo{journal}{J. Phys. Soc. Jpn.} \textbf{\bibinfo{volume}{76}},
  \bibinfo{pages}{073702} (\bibinfo{year}{2007}).

\bibitem[{\citenamefont{Oh et~al.}(2011)\citenamefont{Oh, Jeon, Haam, Park,
  Correa, Lacerda, Cheong, Jeon, and Kim}}]{Oh11}
\bibinfo{author}{\bibfnamefont{Y.S.}~\bibnamefont{Oh}},
  \bibinfo{author}{\bibfnamefont{B.G.} \bibnamefont{Jeon}},
  \bibinfo{author}{\bibfnamefont{S.Y.}~\bibnamefont{Haam}},
  \bibinfo{author}{\bibfnamefont{S.}~\bibnamefont{Park}},
  \bibinfo{author}{\bibfnamefont{V.F.}~\bibnamefont{Correa}},
  \bibinfo{author}{\bibfnamefont{A.H.}~\bibnamefont{Lacerda}},
  \bibinfo{author}{\bibfnamefont{S.W.} \bibnamefont{Cheong}},
  \bibinfo{author}{\bibfnamefont{G.S.}~\bibnamefont{Jeon}}, \bibnamefont{and}
  \bibinfo{author}{\bibfnamefont{K.H.}~\bibnamefont{Kim}},
  \bibinfo{title}{\textit{Strong magnetoelastic effect on the magnetoelectric phenomena of TbMn$_2$O$_5$}},
  \bibinfo{journal}{Phys. Rev. B} \textbf{\bibinfo{volume}{83}},
  \bibinfo{pages}{060405} (\bibinfo{year}{2011}).

\bibitem[{\citenamefont{Johnson et~al.}(2011)\citenamefont{Johnson, Mazzoli,
  Bland, Du, and Hatton}}]{Johnson11}
\bibinfo{author}{\bibfnamefont{R.D.}~\bibnamefont{Johnson}},
  \bibinfo{author}{\bibfnamefont{C.}~\bibnamefont{Mazzoli}},
  \bibinfo{author}{\bibfnamefont{S.R.}~\bibnamefont{Bland}},
  \bibinfo{author}{\bibfnamefont{C.H.} \bibnamefont{Du}}, \bibnamefont{and}
  \bibinfo{author}{\bibfnamefont{P.D.}~\bibnamefont{Hatton}},
  \bibinfo{title}{\textit{Magnetically induced electric polarization reversal in multiferroic TbMn$_2$O$_5$: Terbium spin reorientation studied by resonant x-ray diffraction}},
  \bibinfo{journal}{Phys. Rev. B} \textbf{\bibinfo{volume}{83}},
  \bibinfo{pages}{054438} (\bibinfo{year}{2011}).

\bibitem[{\citenamefont{Lottermoser et~al.}(2009)\citenamefont{Lottermoser,
  Meier, Pisarev, and Fiebig}}]{Lottermoser09}
\bibinfo{author}{\bibfnamefont{Th.}~\bibnamefont{Lottermoser}},
  \bibinfo{author}{\bibfnamefont{D.}~\bibnamefont{Meier}},
  \bibinfo{author}{\bibfnamefont{R.V.}~\bibnamefont{Pisarev}}, \bibnamefont{and}
  \bibinfo{author}{\bibfnamefont{M.}~\bibnamefont{Fiebig}},
  \bibinfo{title}{\textit{Giant coupling of second-harmonic generation to multiferroic polarization}},
  \bibinfo{journal}{Phys. Rev. B} \textbf{\bibinfo{volume}{80}},
  \bibinfo{pages}{100101(R)} (\bibinfo{year}{2009}).

\bibitem[{\citenamefont{Tol{\'e}dano et~al.}(2009)\citenamefont{Tol{\'e}dano,
  Schranz, and Krexner}}]{Toledano09}
\bibinfo{author}{\bibfnamefont{P.}~\bibnamefont{Tol{\'e}dano}},
  \bibinfo{author}{\bibfnamefont{W.}~\bibnamefont{Schranz}}, \bibnamefont{and}
  \bibinfo{author}{\bibfnamefont{G.}~\bibnamefont{Krexner}},
  \bibinfo{title}{\textit{Induced ferroelectric phases in TbMn$_2$O$_5$}},
  \bibinfo{journal}{Phys. Rev. B} \textbf{\bibinfo{volume}{79}},
  \bibinfo{pages}{144103} (\bibinfo{year}{2009}).

\bibitem[{\citenamefont{{Men'shenin}}(2009)}]{Menshenin09}
\bibinfo{author}{\bibfnamefont{V.V.}~\bibnamefont{{Menshenin}}},
  \bibinfo{title}{\textit{Interrelation between the soliton lattice and electric polarization in RMn$_2$O$_5$ oxides}},
  \bibinfo{journal}{JETP}
  \textbf{\bibinfo{volume}{108}}, \bibinfo{pages}{236} (\bibinfo{year}{2009}).

\bibitem[{\citenamefont{Toledano and Toledano}(1987)}]{Toledano87}
\bibinfo{author}{\bibfnamefont{J.-C.} \bibnamefont{Toledano}} \bibnamefont{and}
  \bibinfo{author}{\bibfnamefont{P.}~\bibnamefont{Toledano}},
  \emph{\bibinfo{title}{The Landau theory of phase transitions}}
  (\bibinfo{publisher}{World Scientific Publishing Co. Pte. Ltd.},
  \bibinfo{year}{1987}).

\bibitem[{\citenamefont{Kagomiya et~al.}(2002)\citenamefont{Kagomiya, Kohn, and
  Uchiyama}}]{Kagomiya02}
\bibinfo{author}{\bibfnamefont{I.}~\bibnamefont{Kagomiya}},
  \bibinfo{author}{\bibfnamefont{K.}~\bibnamefont{Kohn}}, \bibnamefont{and}
  \bibinfo{author}{\bibfnamefont{T.}~\bibnamefont{Uchiyama}},
  \bibinfo{title}{\textit{Structure and Ferroelectricity of RMn$_2$O$_5$}},
  \bibinfo{journal}{Ferroelectrics} \textbf{\bibinfo{volume}{280}},
  \bibinfo{pages}{131} (\bibinfo{year}{2002}).

\bibitem[{\citenamefont{Kobayashi et~al.}(2004)\citenamefont{Kobayashi, Osawa,
  Kimura, Noda, Kasahara, Mitsuda, and Kohn}}]{Kobayashi04Tb}
\bibinfo{author}{\bibfnamefont{S.}~\bibnamefont{Kobayashi}},
  \bibinfo{author}{\bibfnamefont{T.}~\bibnamefont{Osawa}},
  \bibinfo{author}{\bibfnamefont{H.}~\bibnamefont{Kimura}},
  \bibinfo{author}{\bibfnamefont{Y.}~\bibnamefont{Noda}},
  \bibinfo{author}{\bibfnamefont{N.}~\bibnamefont{Kasahara}},
  \bibinfo{author}{\bibfnamefont{S.}~\bibnamefont{Mitsuda}}, \bibnamefont{and}
  \bibinfo{author}{\bibfnamefont{K.}~\bibnamefont{Kohn}},
  \bibinfo{title}{\textit{Neutron Diffraction Study of Successive Magnetic Phase Transitions in Ferroelectric TbMn$_2$O$_5$}},
  \bibinfo{journal}{J. Phys. Soc. Jpn.} \textbf{\bibinfo{volume}{73}}, \bibinfo{pages}{3439}
  (\bibinfo{year}{2004}).

\bibitem[{\citenamefont{Johnson et~al.}(2008)\citenamefont{Johnson, Bland,
  Mazzoli, Beale, Du, Detlefs, Wilkins, and Hatton}}]{Johnson08}
\bibinfo{author}{\bibfnamefont{R.D.}~\bibnamefont{Johnson}},
  \bibinfo{author}{\bibfnamefont{S.R.}~\bibnamefont{Bland}},
  \bibinfo{author}{\bibfnamefont{C.}~\bibnamefont{Mazzoli}},
  \bibinfo{author}{\bibfnamefont{T.A.W.}~\bibnamefont{Beale}},
  \bibinfo{author}{\bibfnamefont{C.H.} \bibnamefont{Du}},
  \bibinfo{author}{\bibfnamefont{C.}~\bibnamefont{Detlefs}},
  \bibinfo{author}{\bibfnamefont{S.B.}~\bibnamefont{Wilkins}}, \bibnamefont{and}
  \bibinfo{author}{\bibfnamefont{P.D.}~\bibnamefont{Hatton}},
  \bibinfo{title}{\textit{Determination of magnetic order of the rare-earth ions in multiferroic TbMn$_2$O$_5$}},
  \bibinfo{journal}{Phys. Rev. B} \textbf{\bibinfo{volume}{78}},
  \bibinfo{pages}{104407} (\bibinfo{year}{2008}).

\bibitem[{\citenamefont{Shen}(2002)}]{Shen_nlo}
\bibinfo{author}{\bibfnamefont{Y.}~\bibnamefont{Shen}},
  \emph{\bibinfo{title}{The Principles of Nonlinear Optics}}
  (\bibinfo{publisher}{John Wiley \& Sons}, \bibinfo{year}{2002}).

\bibitem[{\citenamefont{Fiebig et~al.}(2005)\citenamefont{Fiebig, Pavlov, and
  Pisarev}}]{Fiebig05b}
\bibinfo{author}{\bibfnamefont{M.}~\bibnamefont{Fiebig}},
  \bibinfo{author}{\bibfnamefont{V.V.}~\bibnamefont{Pavlov}}, \bibnamefont{and}
  \bibinfo{author}{\bibfnamefont{R.V.}~\bibnamefont{Pisarev}},
  \bibinfo{title}{\textit{Second-harmonic generation as a tool for studying electronic and magnetic crystals: review}},
  \bibinfo{journal}{J. Opt. Soc. Am. B} \textbf{\bibinfo{volume}{22}},
  \bibinfo{pages}{96} (\bibinfo{year}{2005}).

\bibitem[{\citenamefont{Uesu et~al.}(1995)\citenamefont{Uesu, Kurimura, and
  Yamamoto}}]{Uesu95}
\bibinfo{author}{\bibfnamefont{Y.}~\bibnamefont{Uesu}},
  \bibinfo{author}{\bibfnamefont{S.}~\bibnamefont{Kurimura}}, \bibnamefont{and}
  \bibinfo{author}{\bibfnamefont{Y.}~\bibnamefont{Yamamoto}},
  \bibinfo{title}{\textit{New nonlinear optical microscope and its application to the observation of ferroelectric domain structure}},
  \bibinfo{journal}{Ferroelectrics} \textbf{\bibinfo{volume}{169}},
  \bibinfo{pages}{249} (\bibinfo{year}{1995}).

\bibitem[{\citenamefont{Cruz et~al.}(2007)\citenamefont{Cruz, Lorenz, Sun,
  Wang, Park, Cheong, Gospodinov, and Chu}}]{Cruz07}
\bibinfo{author}{\bibfnamefont{C.R.}~\bibnamefont{dela Cruz}},
  \bibinfo{author}{\bibfnamefont{B.}~\bibnamefont{Lorenz}},
  \bibinfo{author}{\bibfnamefont{Y.Y.}~\bibnamefont{Sun}},
  \bibinfo{author}{\bibfnamefont{Y.}~\bibnamefont{Wang}},
  \bibinfo{author}{\bibfnamefont{S.}~\bibnamefont{Park}},
  \bibinfo{author}{\bibfnamefont{S.W.} \bibnamefont{Cheong}},
  \bibinfo{author}{\bibfnamefont{M.M.}~\bibnamefont{Gospodinov}},
  \bibnamefont{and} \bibinfo{author}{\bibfnamefont{C.W.}~\bibnamefont{Chu}},
  \bibinfo{title}{\textit{Pressure-induced enhancement in multiferroic $R$Mn$_2$O$_5$ ($R$=Tb,Dy,Ho)}},
  \bibinfo{journal}{Phys. Rev. B} \textbf{\bibinfo{volume}{76}},
  \bibinfo{pages}{174106} (\bibinfo{year}{2007}).

\bibitem[{\citenamefont{Radaelli and Chapon}(2008)}]{Radaelli08}
\bibinfo{author}{\bibfnamefont{P.}~\bibnamefont{Radaelli}} \bibnamefont{and}
  \bibinfo{author}{\bibfnamefont{L.}~\bibnamefont{Chapon}},
  \bibinfo{title}{\textit{A neutron diffraction study of RMn$_2$O$_5$ multiferroics}},
  \bibinfo{journal}{J. Phys.: Condens. Matter} \textbf{\bibinfo{volume}{20}},
  \bibinfo{pages}{434213} (\bibinfo{year}{2008}).

\bibitem[{\citenamefont{Fukunaga and Noda}(2010)}]{Fukunaga10}
\bibinfo{author}{\bibfnamefont{M.}~\bibnamefont{Fukunaga}} \bibnamefont{and}
  \bibinfo{author}{\bibfnamefont{Y.}~\bibnamefont{Noda}},
  \bibinfo{title}{\textit{Classification and Interpretation of the Polarization of Multiferroic \textit{R}Mn$_2$O$_5$}},
  \bibinfo{journal}{J. Phys. Soc. Jpn.} \textbf{\bibinfo{volume}{79}}, \bibinfo{pages}{054705}
  (\bibinfo{year}{2010}).

\bibitem[{\citenamefont{Wehrenfennig et~al.}(2010)\citenamefont{Wehrenfennig,
  Meier, Lottermoser, Lonkai, Hoffmann, Aliouane, Argyriou, and
  Fiebig}}]{Wehrenfennig10}
\bibinfo{author}{\bibfnamefont{C.}~\bibnamefont{Wehrenfennig}},
  \bibinfo{author}{\bibfnamefont{D.}~\bibnamefont{Meier}},
  \bibinfo{author}{\bibfnamefont{Th.}~\bibnamefont{Lottermoser}},
  \bibinfo{author}{\bibfnamefont{T.}~\bibnamefont{Lonkai}},
  \bibinfo{author}{\bibfnamefont{J.-U.} \bibnamefont{Hoffmann}},
  \bibinfo{author}{\bibfnamefont{N.}~\bibnamefont{Aliouane}},
  \bibinfo{author}{\bibfnamefont{D.~N.} \bibnamefont{Argyriou}},
  \bibnamefont{and} \bibinfo{author}{\bibfnamefont{M.}~\bibnamefont{Fiebig}},
  \bibinfo{title}{\textit{Incompatible magnetic order in multiferroic hexagonal DyMnO$_3$}},
  \bibinfo{journal}{Phys. Rev. B} \textbf{\bibinfo{volume}{82}},
  \bibinfo{pages}{100414} (\bibinfo{year}{2010}).

\bibitem[{\citenamefont{Blake et~al.}(2005)\citenamefont{Blake, Chapon,
  Radaelli, Park, Hur, Cheong, and Rodr\'{\i}guez-Carvajal}}]{Blake05}
\bibinfo{author}{\bibfnamefont{G.R.}~\bibnamefont{Blake}},
  \bibinfo{author}{\bibfnamefont{L.C.}~\bibnamefont{Chapon}},
  \bibinfo{author}{\bibfnamefont{P.G.}~\bibnamefont{Radaelli}},
  \bibinfo{author}{\bibfnamefont{S.}~\bibnamefont{Park}},
  \bibinfo{author}{\bibfnamefont{N.}~\bibnamefont{Hur}},
  \bibinfo{author}{\bibfnamefont{S.W.} \bibnamefont{Cheong}},
  \bibnamefont{and}
  \bibinfo{author}{\bibfnamefont{J.}~\bibnamefont{Rodr\'{\i}guez-Carvajal}},
  \bibinfo{title}{\textit{Spin structure and frustration in multiferroic $R$Mn$_2$O$_5$ ($R$ = Tb,Ho,Dy)}},
  \bibinfo{journal}{Phys. Rev. B} \textbf{\bibinfo{volume}{71}},
  \bibinfo{pages}{214402} (\bibinfo{year}{2005}).

\bibitem[{\citenamefont{Koo et~al.}(2007)\citenamefont{Koo, Song, Ji, Lee,
  Park, Jang, Yang, Park, Jeong, Lee et~al.}}]{Koo07}
\bibinfo{author}{\bibfnamefont{J.}~\bibnamefont{Koo}},
  \bibinfo{author}{\bibfnamefont{C.}~\bibnamefont{Song}},
  \bibinfo{author}{\bibfnamefont{S.}~\bibnamefont{Ji}},
  \bibinfo{author}{\bibfnamefont{J.-S.} \bibnamefont{Lee}},
  \bibinfo{author}{\bibfnamefont{J.}~\bibnamefont{Park}},
  \bibinfo{author}{\bibfnamefont{T.-H.} \bibnamefont{Jang}},
  \bibinfo{author}{\bibfnamefont{C.-H.} \bibnamefont{Yang}},
  \bibinfo{author}{\bibfnamefont{J.-H.} \bibnamefont{Park}},
  \bibinfo{author}{\bibfnamefont{Y.}~\bibnamefont{Jeong}},
  \bibinfo{author}{\bibfnamefont{K.-B.} \bibnamefont{Lee}},
  \bibinfo{title}{\textit{Non-Resonant and Resonant X-Ray Scattering Studies on Multiferroic TbMn$_2$O$_5$}},
  \bibnamefont{et~al.}, \bibinfo{journal}{Phys. Rev. Lett.}
  \textbf{\bibinfo{volume}{99}}, \bibinfo{pages}{197601}
  (\bibinfo{year}{2007}).

\bibitem[{\citenamefont{Chapon et~al.}(2004)\citenamefont{Chapon, Blake,
  Gutmann, Park, Hur, Radaelli, and Cheong}}]{Chapon04}
\bibinfo{author}{\bibfnamefont{L.C.}~\bibnamefont{Chapon}},
  \bibinfo{author}{\bibfnamefont{G.R.}~\bibnamefont{Blake}},
  \bibinfo{author}{\bibfnamefont{M.J.}~\bibnamefont{Gutmann}},
  \bibinfo{author}{\bibfnamefont{S.}~\bibnamefont{Park}},
  \bibinfo{author}{\bibfnamefont{N.}~\bibnamefont{Hur}},
  \bibinfo{author}{\bibfnamefont{P.G.}~\bibnamefont{Radaelli}}, \bibnamefont{and}
  \bibinfo{author}{\bibfnamefont{S.W.} \bibnamefont{Cheong}},
  \bibinfo{title}{\textit{Structural Anomalies and Multiferroic Behavior in Magnetically Frustrated TbMn$_2$O$_5$}},
  \bibinfo{journal}{Phys. Rev. Lett.} \textbf{\bibinfo{volume}{93}},
  \bibinfo{pages}{177402} (\bibinfo{year}{2004}).

\bibitem[{\citenamefont{Saito and Kohn}(1995)}]{Saito95}
\bibinfo{author}{\bibfnamefont{K.}~\bibnamefont{Saito}} \bibnamefont{and}
  \bibinfo{author}{\bibfnamefont{K.}~\bibnamefont{Kohn}},
  \bibinfo{title}{\textit{Magnetoelectric effect and low-temperature phase transitions of TbMn$_2$O$_5$}},
  \bibinfo{journal}{J. Phys. Soc. Jpn.} \textbf{\bibinfo{volume}{7}}, \bibinfo{pages}{2855}
  (\bibinfo{year}{1995}).

\bibitem[{\citenamefont{Nakamura and Kohn}(1997)}]{Nakamura97}
\bibinfo{author}{\bibfnamefont{H.}~\bibnamefont{Nakamura}} \bibnamefont{and}
  \bibinfo{author}{\bibfnamefont{K.}~\bibnamefont{Kohn}},
  \bibinfo{title}{\textit{Magnetoelectric Effect of Rare Earth Manganese Oxide RMn$_2$O$_5$}},
  \bibinfo{journal}{Ferroelectrics} \textbf{\bibinfo{volume}{204}},
  \bibinfo{pages}{107} (\bibinfo{year}{1997}).

\bibitem[{\citenamefont{Moskvin and Pisarev}(2008)}]{Moskvin08Tb}
\bibinfo{author}{\bibfnamefont{A.S.}~\bibnamefont{Moskvin}} \bibnamefont{and}
  \bibinfo{author}{\bibfnamefont{R.V.}~\bibnamefont{Pisarev}},
  \bibinfo{title}{\textit{Charge-transfer in mixed-valent multiferroic TbMn$_2$O$_5$}},
  \bibinfo{journal}{Phys. Rev. B} \textbf{\bibinfo{volume}{77}},
  \bibinfo{pages}{060102(R)} (\bibinfo{year}{2008}).

\bibitem[{\citenamefont{Beale et~al.}(2010)\citenamefont{Beale, Wilkins,
  Johnson, Bland, Joly, Forrest, McMorrow, Yakhou, Prabhakaran, Boothroyd
  et~al.}}]{Beale10}
\bibinfo{author}{\bibfnamefont{T.}~\bibnamefont{Beale}},
  \bibinfo{author}{\bibfnamefont{S.}~\bibnamefont{Wilkins}},
  \bibinfo{author}{\bibfnamefont{R.}~\bibnamefont{Johnson}},
  \bibinfo{author}{\bibfnamefont{S.}~\bibnamefont{Bland}},
  \bibinfo{author}{\bibfnamefont{Y.}~\bibnamefont{Joly}},
  \bibinfo{author}{\bibfnamefont{T.}~\bibnamefont{Forrest}},
  \bibinfo{author}{\bibfnamefont{D.}~\bibnamefont{McMorrow}},
  \bibinfo{author}{\bibfnamefont{F.}~\bibnamefont{Yakhou}},
  \bibinfo{author}{\bibfnamefont{D.}~\bibnamefont{Prabhakaran}},
  \bibinfo{author}{\bibfnamefont{A.}~\bibnamefont{Boothroyd}},
  \bibinfo{title}{\textit{Antiferromagnetically spin polarized oxygen in magneto-electric TbMn$_2$O$_5$}},
  \bibnamefont{et~al.}, \bibinfo{journal}{Phys. Rev. Lett.}
  \textbf{\bibinfo{volume}{105}}, \bibinfo{pages}{087203}
  (\bibinfo{year}{2010}).

\bibitem[{\citenamefont{Partzsch et~al.}(2011)\citenamefont{Partzsch, Wilkins,
  Hill, Schierle, Weschke, Souptel, B\"uchner, and Geck}}]{Partzsch11}
\bibinfo{author}{\bibfnamefont{S.}~\bibnamefont{Partzsch}},
  \bibinfo{author}{\bibfnamefont{S.~B.} \bibnamefont{Wilkins}},
  \bibinfo{author}{\bibfnamefont{J.~P.} \bibnamefont{Hill}},
  \bibinfo{author}{\bibfnamefont{E.}~\bibnamefont{Schierle}},
  \bibinfo{author}{\bibfnamefont{E.}~\bibnamefont{Weschke}},
  \bibinfo{author}{\bibfnamefont{D.}~\bibnamefont{Souptel}},
  \bibinfo{author}{\bibfnamefont{B.}~\bibnamefont{B\"uchner}},
  \bibnamefont{and} \bibinfo{author}{\bibfnamefont{J.}~\bibnamefont{Geck}},
  \bibinfo{title}{\textit{Observation of Electronic Ferroelectric Polarization in Multiferroic YMn$_2$O$_5$}},
  \bibinfo{journal}{Phys. Rev. Lett.} \textbf{\bibinfo{volume}{107}},
  \bibinfo{pages}{057201} (\bibinfo{year}{2011}).

\end{thebibliography}

\end{document}